\documentclass[10pt,english]{article}
\usepackage[T1]{fontenc}
\usepackage[latin9]{inputenc}
\usepackage{geometry}
\usepackage{array}
\usepackage{textcomp}
\usepackage{amsmath}
\usepackage{amssymb}

\makeatletter

\providecommand{\tabularnewline}{\\}

\newcommand{\lyxaddress}[1]{
\par {\raggedright #1
\vspace{1.4em}
\noindent\par}
}
\newenvironment{lyxlist}[1]
{\begin{list}{}
{\settowidth{\labelwidth}{#1}
 \setlength{\leftmargin}{\labelwidth}
 \addtolength{\leftmargin}{\labelsep}
 }}
{\end{list}}

\usepackage{babel}

\makeatother

\usepackage{babel}
\begin{document}

\title{Interpolation Methods for Binary and Multivalued Logical Quantum
Gate Synthesis}

\author{Zeno TOFFANO\textbf{$^{a,b}$}, François DUBOIS\textbf{$^{c,d}$}}
\maketitle

\lyxaddress{\begin{center}
\textit{\small{}$^{a}$ Telecom Dept., CentraleSupélec, Gif-sur-Yvette,
France}{\small{}}\\
{\small{} }\textit{\small{}$^{b}$ Lab. des Signaux et Systèmes -
L2S (UMR8506) - CNRS - U. Paris-Saclay, France}{\small{}}\\
{\small{} }\textit{\small{}$^{c}$ LMSSC, CNAM, Paris, France}{\small{}}\\
{\small{} }\textit{\small{}$^{d}$ Department of Mathematics, University
Paris-Sud, Orsay, France}{\small{}}\\
{\small{} zeno.toffano@centralesupelec.fr ; francois.dubois@math.u-psud.fr
} 
\par\end{center}}
\begin{abstract}
A method for synthesizing quantum gates is presented based on interpolation
methods applied to operators in Hilbert space. Starting from the diagonal
forms of specific generating seed operators with non-degenerate eigenvalue
spectrum one obtains for arity-one a complete family of logical operators
corresponding to all the one-argument logical connectives. Scaling-up
to $n$-arity gates is obtained by using the Kronecker product and
unitary transformations. The quantum version of the Fourier transform
of Boolean functions is presented and a Reed-Muller decomposition
for quantum logical gates is derived. The common control gates can
be easily obtained by considering the logical correspondence between
the control logic operator and the binary propositional logic operator.
A new polynomial and exponential formulation of the Toffoli gate is
presented. The method has parallels to quantum gate\textit{-T} optimization
methods using powers of multilinear operator polynomials. The method
is then applied naturally to alphabets greater than two for multi-valued
logical gates used for quantum Fourier transform, min-max decision
circuits and multivalued adders.

{[}development of the accepted poster presentation at the ``Theory
of Quantum Computation, Communication and Cryptography'' Conference
TQC2017, Paris, France, June 14-16, 2017{]}.

\end{abstract}

\section{Introduction}

Quantum calculation methods are becoming a strategic issue for emerging
technologies such as quantum computing and quantum simulation. Because
of the limited (at this date) quantum computational resources available
it is very important to reduce the resources required to implement
a given quantum circuit. The problem of optimal quantum circuit synthesis
is an important topic much of the effort focuses on decomposition
methods of logical gates. Quantum reversible gates have been extensively
studied using principally Boolean functions implemented on Clifford,
controlled-not and Toffoli gates and also non-Clifford such as the
$T$ gate. The advantages and drawbacks for privileging certain families
of gates has been thoroughly investigated. Basically all methods rely
on the control-logic paradigm as was first proposed by David Deutsch
in \cite{key-1}. The $2$-qubit controlled-not gate being its simplest
element resulting in the operation $|x,y>\rightarrow|x,x\oplus y>$
where the exclusive disjunction ($XOR$, $\oplus$) acts as logical
negation on qubit $y$ when the control qubit $x$ is at one and leaves
it unchanged otherwise. The universality of this logic is assured
by the double-controlled-not or Toffoli gate with the operation on
three qubits $|x,y,z>\rightarrow|x,y,xy\oplus z>$. This gate operates
on the third qubit, $z$, when the conjunction is met on the control
bits $x$ and $y$ (both must be $1$) being thus equivalent to a
negated binary conjunction $NAND$ gate, which is known to be universal.
These logical controlled-gates transform the qubits as reversible
permutation operators \textit{i.e.} they are not diagonal in the computational
basis.

In recent years many synthesizing methods for quantum circuits actually
use operators that are diagonal in the computational basis because
of the simpler resulting mathematical operations. The controlled-Z
and the double-controlled-$Z$ gates are used to build controlled-not
and Toffoli gates. Solutions using diagonal $T$ and $S$ gates, as
will be shown hereafter, are also popular, also the stabilizer formalism
and surface codes use families of Clifford diagonal gates. So in some
way the ``back to diagonal'' trend seems to present benefits, this
is also justified when one considers questions in quantum physics
where much effort is employed to define Hamiltonians and the corresponding
energy spectrum (eigenvalues) and stationary states (eigenstates),
which can then be implemented in quantum simulation models.

A question arises: can logical calculations be formalized someway
directly in the qubit eigenspace? The answer seems to be affirmative
and in a rather simple way too. A proposal has been given by making
parallels between propositional logic and operator linear algebra
in the framework of ``Eigenlogic'' \cite{key-2,key-3}, the idea
is that Boolean algebra operations can be represented by operators
in Hilbert space, using projection operators as developed in \cite{key-2}
and extended to angular momentum operators as shown in \cite{key-3}
and generalized in this paper.

A theoretical justification could be inspired by Pierre Cartier in
\cite{key-30}, relating the link between the algebra of logical propositions
and the set of all valuations on it, he writes: ``...in the \textit{theory
of models} in logic a model of a set of propositions has the effect
of validating certain propositions. With each logical proposition
one can associate by duality the set of all its true valuations represented
by the number $1$. This correspondence makes it possible to interpret
the algebra of propositions as a class of subsets, conjunction and
disjunction becoming respectively the intersection and union of sets.
This corresponds to the \textit{Stone duality} proved by the Stone
representation theorem\textit{ }and is one of the spectacular successes
of twentieth century mathematics....The development of quantum theory
led to the concept of a quantum state, which can be understood as
a new embodiment of the concept of a valuation''. The idea is not
new, and stems from John Von Neumann's proposal of ``projections
as propositions'' in \cite{key-31} which was subsequently formalized
in quantum logic with Garret Birkhoff in \cite{key-32}. These topics
have been thoroughly discussed in \cite{key-2}.

But because nowadays quantum logic is mostly interested in aspects
concerning operations going beyond and also in contrast with the principles
of classical logic it is still not considered as an operational tool
for quantum computing even though many bridges have been made \cite{key-33,key-34}.
The aim of Eigenlogic is on the other side to fully exploit the logical
structure offered by the operational system in the eigenspace with
of course the possibility to look outside at other basis representations
this has led for example to \textit{fuzzy logic} applications in \cite{key-3}.

While most of the research is currently devoted to quantum circuit
and algorithm developments based on the use of quantum operators working
in binary (Boolean) systems through the manipulation of qubits, there
are many possibilities for the exploitation of observables with more
than two non-degenerate eigenvalues and the theory of multi-valued-logic
is very naturally applicable to the design and analysis of these systems.

\section{The interpolation method for quantum operators}

Generalizing the method used in \cite{key-2,key-3} we look for $n$-arity
logical operators supporting a finite set of $m$ distinct logical
values.

\subsection{The Seed Operator and the associated projection operators}

The method presented here is inspired from classical \textit{La}z\textit{grange
interpolation} where the ``variable'' is represented by a \textit{seed
operator} acting in Hilbert space, possessing $m$ distinct eigenvalues,
\textit{i.e.} it is non-degenerate. The values are not fixed meaning
that one can work with different alphabets, in the binary case one
uses classically the Boolean values $\left\{ 0,1\right\} $, but also
$\left\{ +1,-1\right\} $ are often considered. What this method will
show is that for whatever finite system of values, unique logical
operators can be defined. In the multivalued case, popular choices
are, for example, the system $\left\{ 0,1,...,m-1\right\} $ formalized
in Post logic \cite{key-4} and rational fractional values in the
unit interval $\left[0,1\right]$ giving the system$\left\{ 0,\frac{1}{m},\frac{2}{m}...,\frac{m-1}{m},1\right\} $
as used in \L ukasiewicz logic. Other convenient numerical choices
are the balanced ternary values $\left\{ +1,0,-1\right\} $, these
can characterize qutrits associated to quantum orbital angular momentum
observables (see \cite{key-3} and hereafter).

When the eigenvalues are real the corresponding logical operators
are \textit{observables}, \textit{e.g.} Hermitian operators. Many
logical gates are of this type, the most popular ones are the one-qubit
Pauli-$Z$, Pauli-$X$, Hadamard $H$, phase $S$ and $T$ gates,
the two-qubit controlled-not, controlled-Z gate and the swap gate
and the universal three-qubit Toffoli gate. These observable gates
are also named classical gates.

The possibility of using complex values can also be considered and
this has applications for specific problems such as in $T$-gate synthesis
and quantum Fourier transform in this last case one considers the
roots of unity for $m$ values giving the spectrum $\left\{ e^{i2\pi\cdot0},e^{i2\pi\cdot\frac{1}{m}},e^{i2\pi\cdot\frac{2}{m}},...,e^{i2\pi\cdot\frac{m-1}{m}}\right\} $.

In each situation one starts by defining the \textit{seed operator}
$\boldsymbol{\Lambda}$ with $m$ non-degenerate eigenvalues $\lambda_{i}$.
The density matrix of eigenstate $|\lambda_{i}>$ is $\boldsymbol{\Pi}_{\lambda_{i}}=|\lambda_{i}><\lambda_{i}|$,
and corresponds to the rank-1 projection operator,\textit{ a ray},
for this state. It is obtained by the following expression:

\begin{equation}
\boldsymbol{\Pi}_{\lambda_{i}}(\boldsymbol{\Lambda})=\prod_{j=1\,(j\neq i)}^{m}\frac{(\boldsymbol{\Lambda}-\lambda_{i}\mathbf{I}_{m})}{(\lambda_{i}-\lambda_{j})}\label{eq:Lagr_projector}
\end{equation}
this expression corresponds to the \textit{Duncan-Collar formula.}
The development is unique according to the \textit{Cayley-Hamilton
theorem}. Expression (\ref{eq:Lagr_projector}) is polynomial consisting
in a linear combination of powers up to $m-1$ of the seed operator
$\boldsymbol{\Lambda}$, it is a rank-1 projection operator over the
$m$ dimensional Hilbert space and can be expressed as a $m\times m$
square matrix.

The number $m$ of rays is the dimension of the vector space. All
the rank-$1$ projection operators obtained in (\ref{eq:Lagr_projector})
for the eigensystem commute and span the entire space by the closure
relation:

\begin{equation}
{\displaystyle \sum_{i=1}^{m}\boldsymbol{\Pi}_{\lambda_{i}}=\sum_{i=1}^{m}|\lambda_{i}><\lambda_{i}|}=\mathbf{I}_{m}\label{eq:Closure-Projectors}
\end{equation}

\subsection{Logical operators for arity-one}

As outlined in \cite{key-2,key-3} in Eigenlogic eigenvalues correspond
to logical truth values (values $\left\{ 0,1\right\} $ for a Boolean
system) and one can make the correspondence between propositional
logical connectives and operators in Hilbert space. The truth table
of a logical connective corresponds to different semantic \textit{interpretation}s,
each interpretation is a fixed attribution of truth values to the
elementary propositions (the ``inputs'') composing the connective.
In Eigenlogic each interpretation corresponds to one of the eigenvectors
and the associated eigenvalue to the corresponding truth-value for
the considered logical connective.

It is well known that logical connectives can be expressed through
arithmetic expressions, these are closely related to polynomial expressions
over rings, but with expressions using ordinary arithmetic addition
and subtraction instead of their modular counterpart as used in Boolean
algebra. These topics were thoroughly discussed in \cite{key-2} for
the Boolean values $\left\{ 0,1\right\} $. Arithmetic developments
of logical connectives are often used, for example, in the description
of switching functions for decision logic design. A good review was
given by Svetlana Yanushkevich in \cite{key-6}.

In logic, the functions and their arguments take the same values and
these are the only possible logical values. For Boolean functions
these unique possible values are the two numbers $0$ and $1$ corresponding
respectively to the \textit{False} and \textit{True} character of
a logical proposition. So considering an arithmetic expression for
an arity-one logical connective $\ell(p)$ and choosing the $m$ distinct
logical values $a_{i}$, the value taken by the logical function at
one of these points $a_{p}$ is $\ell(a_{p})\in\left\{ a_{1},a_{2},...,a_{m}\right\} $,
also one of these values. The corresponding unique logical operator
is given by the interpolation development:

\begin{equation}
\boldsymbol{F}_{\ell}=\sum_{i=1}^{m}\ell(a_{i})\,\boldsymbol{\Pi}_{a_{i}}\label{eq:1-arity-Logc-funct-dev}
\end{equation}
this matrix decomposition formula is proved by \textit{Sylvester's
theorem} and represents the \textit{spectral decomposition} of the
operator. The projection operators are obtained by equation (\ref{eq:Lagr_projector})
with $\lambda_{i}=a_{i}$. It has to be outlined that the method used
for obtaining expression (\ref{eq:1-arity-Logc-funct-dev}) is the
counterpart for operators of the classical Lagrange interpolation
method over $m$ points \cite{key-7}. As for the projectors in (\ref{eq:Lagr_projector})
the logical operator in (\ref{eq:1-arity-Logc-funct-dev}) can also
be put as a linear combination of powers up to $m-1$ of the seed
operator $\boldsymbol{\Lambda}$.

One must then consider the operators corresponding to elementary propositions
of the logical connectives. This is a straightforward procedure in
Eigenlogic. For arity-one logical connectives, these are function
of one single elementary proposition $p$ which corresponds in the
method proposed here to the seed operator $\boldsymbol{\Lambda}$.
These concepts will become clearer with some examples presented hereafter.

\subsection{Scaling up to higher arity}

The scaling is obtained by following the same procedure as for classical
interpolation methods for multivariate systems using tensor products
(see \textit{e.g.} \cite{key-7}). Here the chosen convention for
the indexes is the one given by David Mermin in \cite{key-8} where
index 0 indicates the lowest digit, index 1 the next and so on...
For arity-$2$ one considers two operators $\boldsymbol{P}_{1}$ and
$\boldsymbol{Q}_{0}$ corresponding to the propositional variables
$p$ and $q$. One can write by the means of the Kronecker product
$\otimes$:

\begin{equation}
\boldsymbol{P}_{1}=\boldsymbol{\Lambda}\otimes\mathbf{I}\qquad,\qquad\boldsymbol{Q}_{0}=\mathbf{I}\otimes\boldsymbol{\Lambda}
\end{equation}
for arity-$3$ using three operators one has:

\begin{equation}
\boldsymbol{P}_{2}=\boldsymbol{\Lambda}\otimes\mathbf{I}\otimes\mathbf{I}\:,\qquad\boldsymbol{Q}_{1}=\mathbf{I}\otimes\boldsymbol{\Lambda}\otimes\mathbf{I}\:,\qquad\boldsymbol{R}_{0}=\mathbf{I}\otimes\mathbf{I}\otimes\boldsymbol{\Lambda}
\end{equation}
for higher arity-$n$ the procedure can be automatically iterated.

Logical operators can then be obtained by (\ref{eq:1-arity-Logc-funct-dev}),
and are multi-linear combinations of the elementary operators $\boldsymbol{P}_{i}$,
$\boldsymbol{Q}_{j}$...., for example the simplest operators are
products of these. Due to logical completeness (demonstrated by Emil
Post in \cite{key-4}) there will be for an $m$-valued $n$-arity
system exactly $m^{m^{n}}$ logical operators forming a complete family
of commuting logical operators. Logical completeness has also another
important consequence: there are always universal connectives from
which all the others can be derived, it has been shown, also in \cite{key-4},
that for an $m$-valued arity-$2$ system one needs at least two universal
connectives, in the case of Post logic these turn out to be the \textit{general
negation} (cyclic permutation of all values $a_{i}\rightarrow a_{i+1}$)
and the $Max$ connective (takes the highest of the two input values)
these reduce to Boolean connectives for binary values where the $Max$
connective becomes the disjunction ($OR$, $\vee$). The logical operators
obtained here for an $m$-valued $n$-arity system are represented
by $m^{n}\times m^{n}$ square matrices.

\section{Interpolation synthesis of binary quantum gates}

In quantum computing operations on qubits are done using reversible
gates without reducing the dimension of the vector space, except at
the end of the calculation where storage of some values is done by
projection operators transforming the qubits into classical bits.

\subsection{\label{subsec:Fourier-transform-of}quantum Fourier transform of
Boolean functions }

The following method is based on basic mathematical properties of
logical functions, that have their origin in George Boole's elective
development theorem for logical functions \cite{key-81,key-2}, in
the Fourier analysis of Boolean functions based on the Walsh transform
\cite{key-82,key-83} and on the Reed-Muller developments of Boolean
functions \cite{key-84,key-844}.

Given the \textit{powerset} of a logical system represented by $[n]$
a set of $n$-tuples where each component taking the values either
$0$ or $1$ consists of $2^{n}$ elements, the components range from
$(0,0,...,0)$ to $(1,1,...,1)$. Considering a self-inverse logical
operator $\boldsymbol{G}_{\ell}$ ($\boldsymbol{G}_{\ell}^{2}=\mathbf{I}$)
of arity-$n$ associated to a binary logical function $g_{\ell}:\{+1,\,-1\}^{n}\rightarrow\{+1,\,-1\}$,
where the truth values are $\left\{ +1,-1\right\} $, representing
respectively $False$ and $True$ and corresponding to the Booleans
$\left\{ 0,1\right\} $.

This operator can be expressed as a unique development, as has been
showed here-above, using its spectral decomposition:

\begin{equation}
\boldsymbol{G}_{\ell}=\sum_{\boldsymbol{s}\in[n]}g_{\ell}(\mathbf{s})\,\boldsymbol{\Pi}_{\boldsymbol{s}}\label{eq:Q-Boolean-F-Transf}
\end{equation}
where $\boldsymbol{\Pi}_{\mathbf{s}}$ is a rank-1 projection operator
(of dimension $2^{n}$). One can express this projection operator
using the Pauli $\boldsymbol{Z}=diag(+1,-1)$ gate by:

\begin{equation}
\boldsymbol{\Pi}_{\mathbf{s}}=\bigotimes_{k=1}^{n}\frac{1}{2}(\boldsymbol{\mathrm{I}}+(-1)^{s_{k}}\boldsymbol{Z})\label{eq:Projectors-Z}
\end{equation}
where $\boldsymbol{s}$ represents a $n$-component vector with $s_{k}$
the $k$ component. If $s_{k}=0$ the operator in the bracket gives
the projection operator $\boldsymbol{\Pi}_{0}=\frac{1}{2}(\boldsymbol{\mathrm{I}}+\boldsymbol{Z})$
on state $|0>$ and if $s_{k}=1$ the operator in the bracket is the
projection operator $\boldsymbol{\Pi}_{1}=\frac{1}{2}(\boldsymbol{\mathrm{I}}-\boldsymbol{Z})$
on state $|1>$. $\boldsymbol{\Pi}_{\boldsymbol{s}}$ in equation
(\ref{eq:Projectors-Z}) can then be expanded by means of the following
character operator \cite{key-83}:

\begin{equation}
\boldsymbol{\chi}_{\boldsymbol{p}}=\bigotimes_{r=1}^{n}\boldsymbol{Z}^{p_{r}}\label{eq:character-Z}
\end{equation}
where $\boldsymbol{p}$ represents a $n$-component vector with $p_{r}$
the $r$ component. If $p_{r}=0$ then $\boldsymbol{Z}^{p_{r}=0}=\boldsymbol{\mathrm{I}}$
and one has the identity operator and if $p_{r}=1$ then $\boldsymbol{Z}^{p_{r}=1}=\boldsymbol{Z}$
one has the $\boldsymbol{Z}$-gate. In this way one can re-express
the projection operators of equation (\ref{eq:Projectors-Z}) by:

\begin{equation}
\boldsymbol{\Pi}_{\mathbf{s}}=\frac{1}{2^{n}}\sum_{\boldsymbol{p}\in[n]}(-1)^{\boldsymbol{p}\cdot\boldsymbol{s}}\boldsymbol{\chi}_{\boldsymbol{p}}
\end{equation}
putting this operator back in the expression of $\boldsymbol{G}_{\ell}$
one obtains:

\begin{equation}
\boldsymbol{G}_{\ell}=\frac{1}{2^{n}}\sum_{\boldsymbol{s}\in[n]}g_{\ell}(\mathbf{s})\sum_{\boldsymbol{p}\in[n]}(-1)^{\boldsymbol{p}\cdot\boldsymbol{s}}\boldsymbol{\chi}_{\boldsymbol{p}}=\frac{1}{2^{n}}\sum_{\boldsymbol{p}\in[n]}\left\{ \sum_{\mathbf{s}\in[n]}g_{\ell}(\mathbf{s})(-1)^{\boldsymbol{p}\cdot\boldsymbol{s}}\right\} \boldsymbol{\chi}_{\boldsymbol{p}}
\end{equation}
in this way using the definition of the Boolean Fourier transform
\cite{key-82} :

\begin{equation}
\hat{g}_{\boldsymbol{p}}=\sum_{\boldsymbol{s}\in[n]}(-1)^{\boldsymbol{p}\cdot\boldsymbol{s}}g_{\ell}(\mathbf{s})
\end{equation}
one finally obtains the expression for the self-inverse operator $\boldsymbol{G}_{\ell}$
which can be considered a quantum Boolean transform operator, as defined
in \cite{key-83}:

\begin{equation}
\boldsymbol{G}_{\ell}=\frac{1}{2^{n}}\sum_{\boldsymbol{p}\in[n]}\hat{g}_{\boldsymbol{p}}\boldsymbol{\chi}_{\boldsymbol{p}}\label{eq:QBFT-2}
\end{equation}

\subsection{The Householder transform and Reed-Muller logical decomposition}

There exists a linear bijection between the self-inverse operators
$\boldsymbol{G}$ and the respective idempotent projection operator
$\boldsymbol{\Pi}_{\boldsymbol{G}}$ given by the \textit{Householder
transform}:

\begin{equation}
\boldsymbol{G}=\boldsymbol{\mathrm{I}}-2\boldsymbol{\Pi}_{\boldsymbol{G}}=(-1)^{\boldsymbol{\Pi}_{\boldsymbol{G}}}=e^{i\pi\boldsymbol{\Pi}_{\boldsymbol{G}}}\qquad,\quad\boldsymbol{\Pi}_{\boldsymbol{G}}=\frac{\boldsymbol{\mathrm{I}}-\boldsymbol{G}}{2}\qquad,\quad\boldsymbol{G}=e^{i\frac{\pi}{2}}e^{-i\frac{\pi}{2}\boldsymbol{G}}\label{eq:General-Householder}
\end{equation}

A consequence of equations, (\ref{eq:Q-Boolean-F-Transf}), (\ref{eq:QBFT-2})
and (\ref{eq:General-Householder}) is that all the operators of a
given family $\boldsymbol{G}^{[n]}$ commute and every $\boldsymbol{G}^{[n]}$
commutes with every $\boldsymbol{\Pi}^{[n]}$. This means technically
that all operations on exponential of matrices, as given by equation
(\ref{eq:General-Householder}), are simple arithmetic operations.
For example the product of the exponential of operators will be the
exponential of the sum of the operators. For the logical operators
$\boldsymbol{G}^{[n]}$ the eigenvalues are $+1$ and $-1$, these
numbers represent the truth values of the logical system. Expression
(\ref{eq:Q-Boolean-F-Transf}) can also be derived directly by the
interpolation method and expressed as a function of the corresponding
interpolation projection operators $\boldsymbol{\Pi}_{\boldsymbol{p}}^{[n]}$.
Both expansions in $\boldsymbol{\chi}_{S}^{[n]}$ or $\boldsymbol{\Pi}_{\boldsymbol{p}}^{[n]}$
can be used when transforming operator forms, some examples will be
illustrated hereafter. In general for an $n$-arity system one has
a family of $2^{2^{n}}$ different commuting operators $\boldsymbol{G}_{\ell}^{[n]}$.
Also from the last equation in (\ref{eq:General-Householder}) one
has the simple but important result that the operators $\boldsymbol{G}$
defined as sums in (\ref{eq:Q-Boolean-F-Transf}) can always be tranformed
in products of quantum gates, some examples will be given hereafter.

The Eigenlogic operators \cite{key-3} for exclusive disjunction ($XOR$,
$\oplus$) for an arity-$2$ and arity-$n$ system are defined by:

\begin{equation}
\boldsymbol{G}_{XOR}^{[2]}=diag_{z}(+1,-1,-1,+1)=\boldsymbol{U}_{1}^{[2]}\cdot\boldsymbol{U}_{0}^{[2]}\:,\qquad\boldsymbol{G}_{\oplus^{n}}^{[n]}=\prod_{j=0}^{n-1}\boldsymbol{U}_{j-1}^{[n]}\label{eq:G-XOR-arity2}
\end{equation}
so the function $XOR$ is represented by the matrix product of the
single qubit dictators defined in (\ref{eq:General dictators}) this
operator is local because it is a Kronecker product of local operators.

The corresponding logical projection operator $\boldsymbol{\Pi}_{XOR}$
is given by the polynomial expression (see \cite{key-2,key-6}) using
$\boldsymbol{\Pi}_{0}^{[2]}=\mathbf{I}_{2}\otimes\boldsymbol{\Pi}$
and $\boldsymbol{\Pi}_{1}^{[2]}=\boldsymbol{\Pi}\otimes\mathbf{I}_{2}$
one has :

\begin{equation}
\boldsymbol{\Pi}_{XOR}^{[2]}=\boldsymbol{\Pi}_{1}^{[2]}+\boldsymbol{\Pi}_{0}^{[2]}-2\,\boldsymbol{\Pi}_{1}^{[2]}\cdot\boldsymbol{\Pi}_{0}^{[2]}=diag_{z}(0,1,1,0)
\end{equation}
According to equation (\ref{eq:General-Householder}) this linear
combination becomes the argument of the exponent of $(-1)$ for $\boldsymbol{G}_{XOR}^{[n]}$.
One can thus change all the $-$'s into $+$'s, also all the terms
multiplied by $2^{n}$ with $n>0$ vanish and so:

\begin{equation}
\boldsymbol{G}_{XOR}^{[2]}=\left(-1\right)^{(\boldsymbol{\Pi}_{1}^{[2]}+\boldsymbol{\Pi}_{0}^{[2]}-2\,\boldsymbol{\Pi}_{1}^{[2]}\cdot\boldsymbol{\Pi}_{0}^{[2]})}=\left(-1\right)^{(\boldsymbol{\Pi}_{1}^{[2]}+\boldsymbol{\Pi}_{0}^{[2]})}=\left(-1\right)^{\boldsymbol{\Pi}_{1}^{[2]}}\cdot\left(-1\right)^{\boldsymbol{\Pi}_{0}^{[2]}}=\boldsymbol{U}_{1}^{[2]}\cdot\boldsymbol{U}_{0}^{[2]}
\end{equation}
giving again (\ref{eq:G-XOR-arity2}). Now one can consider explicitly
the action on qubits, with $|x>$, $x\in\left\{ \text{0},1\right\} $,
representing a single qubit eigenstate in the computational basis.
The state for arity-$2$, is $|xy>$ with $x,y\in\left\{ 0,1\right\} $
and so on when scaling for higher arity. The application of the preceding
operators for exclusive disjunction ($XOR$, $\oplus$) on the $2$-qubit
state $|xy>$ and the $3$-qubit state $|xyz>$ gives then:

\begin{equation}
\boldsymbol{G}_{XOR}^{[2]}\,|xy>=\boldsymbol{G}_{x\oplus y}^{[2]}\,|xy>=(-1)^{x+y}|xy>\qquad,\qquad\boldsymbol{G}_{x\oplus y\oplus z}^{[3]}\,|xyz>=(-1)^{x+y+z}|xyz>
\end{equation}

For conjunction ($AND$, $\wedge$) logical operators one has for
a $3$-qubit state $|xyz>$:

\begin{equation}
\boldsymbol{\Pi}_{x\wedge y\wedge z}^{[3]}|xyz>=(xyz)\,|xyz>\qquad,\qquad\boldsymbol{G}_{x\wedge y\wedge z}^{[3]}\,|xyz>=(-1)^{xyz}|xyz>\label{eq:3-G-AND}
\end{equation}

Knowing the polynomial arithmetic expression of the logical connective
one can derive the Reed-Muller form \cite{key-6,key-84,key-844}.
An important example is the disjunction operator ($OR$, $\vee$).
One starts from the arithmetic expansion for the disjunction connective,
which is actually the \textit{inclusion-exclusion} rule, for arity-$3$
this gives:

\begin{equation}
x\vee y\vee z=x+y+z-xy-xz-yz+xyz
\end{equation}
giving

\begin{equation}
\boldsymbol{G}_{x\vee y\vee z}^{[3]}\,|xyz>=(-1)^{x}(-1)^{y}(-1)^{z}(-1)^{xy}(-1)^{xz}(-1)^{yz}(-1)^{xyz}|xyz>
\end{equation}
and the operator form:

\begin{equation}
\boldsymbol{G}_{x\vee y\vee z}^{[3]}=\boldsymbol{U}_{2}^{[3]}\cdot\boldsymbol{U}_{1}^{[3]}\cdot\boldsymbol{U}_{0}^{[3]}\cdot\boldsymbol{G}_{x\wedge y}^{[3]}\cdot\boldsymbol{G}_{x\wedge z}^{[3]}\cdot\boldsymbol{G}_{y\wedge z}^{[3]}\cdot\boldsymbol{G}_{x\wedge y\wedge z}^{[3]}
\end{equation}
which corresponds to the Reed-Muller decomposition of the disjunction
on Booleans:

\begin{equation}
x\vee y\vee z=x\oplus y\oplus z\oplus(x\wedge y)\oplus(x\wedge z)\oplus(y\wedge z)\oplus(x\wedge y\wedge z)
\end{equation}

\subsection{Permutation operators and Pauli gates}

Permutation operators $\boldsymbol{P}$ are unitary operators. Many
quantum gates are permutation operators, \textit{e.g:} Pauli-$\boldsymbol{X}$,
\textit{controlled-not}, \textit{swap} and \textit{Toffoli}, which
are also self-inverse, \textit{e.g.} $\boldsymbol{P}=\boldsymbol{P}^{-1}$
and thus $\boldsymbol{P^{2}}=\mathbf{I}$, meaning that their eigenvalues
are $+1$ and $-1$ and one can apply the method shown above to obtain
their polynomial operator form. Other operators have the same eigenvalues,
this is the case of the Pauli gate group generated by the three Pauli
operators, $\boldsymbol{X}$, $\boldsymbol{Y}$ , $\boldsymbol{Z}$
and the identity operator $\mathbf{I}$ and combined by the Kronecker
product. In the computational basis the $\boldsymbol{Z}$ and $\boldsymbol{X}$
and Hadamard $\boldsymbol{H}$ gate's matrix forms are:

\begin{equation}
\boldsymbol{Z}=\left(\begin{array}{cc}
+1 & 0\\
0 & -1
\end{array}\right)\quad,\qquad\boldsymbol{X}=\left(\begin{array}{cc}
0 & 1\\
1 & 0
\end{array}\right)\quad,\qquad\boldsymbol{H}=\frac{1}{\sqrt{2}}\left(\begin{array}{cc}
+1 & 1\\
1 & -1
\end{array}\right)\label{eq:Z-X-H}
\end{equation}
with the transformation $\boldsymbol{X}=\boldsymbol{H}\cdot\boldsymbol{Z}\cdot\boldsymbol{H}$
and because $\boldsymbol{H}^{2}=\mathbf{I}$ also $\boldsymbol{Z}=\boldsymbol{H}\cdot\boldsymbol{X}\cdot\boldsymbol{H}$:

The seed operator for the system $\left\{ +1,-1\right\} $ in the
computational basis is $\boldsymbol{\Lambda}_{\left\{ +1,-1\right\} }=\boldsymbol{Z}=diag_{z}(+1,-1)$.

Using (\ref{eq:Lagr_projector}) one obtains the two projection operators
on the qubit states $|0>=\left(\begin{array}{c}
1\\
0
\end{array}\right)$ and $|1>=\left(\begin{array}{c}
0\\
1
\end{array}\right)$:

\begin{equation}
\boldsymbol{\Pi}_{+1}(\boldsymbol{Z})=\frac{1}{2}\left(\mathbf{I}+\boldsymbol{Z}\right)=|0><0|\qquad,\quad\boldsymbol{\Pi}_{-1}(\boldsymbol{Z})=\frac{1}{2}\left(\mathbf{I}-\boldsymbol{Z}\right)=|1><1|=\boldsymbol{\Pi}\label{eq:Project-Z}
\end{equation}

The same method can be used when the operators are not diagonal in
the computational basis, for example one could have chosen the $\boldsymbol{X}$
gate as the seed operator leading to similar expressions as in (\ref{eq:Project-Z})
by changing $\boldsymbol{Z}$ into $\boldsymbol{X}$.

The seed operator and the two projection operators permit to write
the four arity-one logical operators: $\boldsymbol{Z}$ (\textit{Dictator}),
$-\boldsymbol{Z}$ (\textit{Negation}), $+\mathbf{I}$ (\textit{Contradiction})
and $-\mathbf{I}$ (\textit{Tautology}) (see \cite{key-2,key-3} for
a detailed discussion), the truth values $\left\{ \text{+1},-1\right\} $
corresponding respectively to \textit{False} and \textit{True}.

Before continuing for higher arity it is interesting to analyze the
logical operator system corresponding to the Boolean values $\left\{ 0,1\right\} $,
it is straightforward to see that in this case the seed operator,
$\boldsymbol{\Lambda}_{\left\{ 0,1\right\} }$, is the projection
operator $\boldsymbol{\Pi}_{1}=\boldsymbol{\Pi}$ given in (\ref{eq:Project-Z}),
the other projection operator $\boldsymbol{\Pi}_{0}$ being the complement:

\begin{equation}
\boldsymbol{\Lambda}_{\left\{ 0,1\right\} }=diag_{z}(0,1)=\boldsymbol{\Pi}_{1}=\boldsymbol{\Pi}=|1><1|\qquad,\qquad\boldsymbol{\Pi}_{0}=\mathbf{I}-\boldsymbol{\Pi}=|0><0|
\end{equation}

As shown above in (\ref{eq:General-Householder}) there are interesting
relations for the operators when going from the system $\left\{ \text{+1},-1\right\} $
to the system $\left\{ 0,1\right\} $. Due to the properties of the
operator $\boldsymbol{\Pi}$ (idempotent projection operator $\boldsymbol{\Pi}^{2}=\boldsymbol{\Pi}$
) and $\boldsymbol{Z}$ (self-inverse operator $\boldsymbol{Z}^{2}=\mathbf{I}$
) one can use the Householder transform:

\begin{equation}
\boldsymbol{Z}=\mathbf{I}-2\boldsymbol{\Pi}=(-1)^{\boldsymbol{\Pi}}=e^{i\pi\boldsymbol{\Pi}}=e^{+i\frac{\pi}{2}}e^{-i\frac{\pi}{2}\boldsymbol{Z}}\quad,\qquad\boldsymbol{Z}\,|x>=(-1)^{x}|x>\label{eq:Householder-Z}
\end{equation}

As shown above in (\ref{eq:General-Householder}) this transform is
also valid for composite expressions and permits to transform operators
with eigenvalues $\left\{ 0,1\right\} $ into operators with eigenvalues
$\left\{ +1,-1\right\} $ and will be used hereafter to build the
controlled-not end Toffoli gates.

For arity-$2$, one considers the two elementary operators acting
on qubit-$0$ and qubit-$1$:

\begin{equation}
\boldsymbol{Z}_{0}=\mathbf{I}\otimes\boldsymbol{Z}\qquad,\quad\boldsymbol{Z}_{1}=\boldsymbol{Z}\otimes\mathbf{I}
\end{equation}

The controlled-z gate, named here $\boldsymbol{C}_{Z}$, is diagonal
in the computational basis and corresponds to a two-argument conjunction
($AND$, $\wedge$) in Eigenlogic. The interpolation method described
above and used in \cite{key-3} gives directly the known \cite{key-8}
polynomial expression:

\begin{equation}
\boldsymbol{C}_{Z}=diag_{z}(+1,+1,+1,-1)=\frac{1}{2}(\mathbf{I}+\boldsymbol{Z}_{1}+\boldsymbol{Z}_{0}-\boldsymbol{Z}_{1}\cdot\boldsymbol{Z}_{0})\label{eq:Cz-polynomial}
\end{equation}

\subsection{Building the controlled-not and Toffoli gates}

One can consider some interesting interpretations using Eigenlogic
\cite{key-3}: the truth table is given by the structure of the eigenvalues
of the $\boldsymbol{C}_{Z}$ operator and corresponds to the logical
connective conjunction ($AND$, $\wedge$), where the values for $\left\{ False,True\right\} $
correspond here to the alphabet $\left\{ +1,-1\right\} $. Also the
logical operator for conjunction $\boldsymbol{\Pi}_{AND}$ in the
Boolean alphabet, $\left\{ x,y\right\} \in\left\{ \text{0},1\right\} $,
as demonstrated in \cite{key-2}, has a very simple form:

\begin{equation}
\boldsymbol{\Pi}_{AND}=\boldsymbol{\Pi}\otimes\boldsymbol{\Pi}=diag_{z}(0,0,0,1)\quad,\qquad\boldsymbol{\Pi}_{AND}\,|xy>=xy\,|xy>
\end{equation}
when applying this operator to the state $|xy>$ it gives the eigenvalue
$xy=1$, $True$, only for the state $|11>$. One can transform this
operator using the Householder transform in:

\begin{equation}
\boldsymbol{Z}_{AND}=\mathbf{I}-2\boldsymbol{\Pi}_{AND}=(-1)^{\boldsymbol{\Pi}\otimes\boldsymbol{\Pi}}=diag_{z}(1,1,1,-1)=\boldsymbol{C}_{Z}\label{eq:Cz-projector}
\end{equation}
\begin{equation}
\boldsymbol{Z}_{AND}\,|xy>=\boldsymbol{C}_{Z}\,|xy>=(-1)^{xy}\,|xy>
\end{equation}
justifying the interpretation of the $\boldsymbol{C}_{Z}$ gate as
a conjunction in Eigenlogic for the alphabet $\left\{ +1,-1\right\} $.

The controlled-not operator, named here $\boldsymbol{C}$, can be
expressed straightforwardly, see \textit{e.g.} \cite{key-8}, in its
polynomial form on $\boldsymbol{Z}_{1}$ and $\boldsymbol{X}_{0}$:

\begin{equation}
\boldsymbol{C}=\frac{1}{2}(\mathbf{I}+\boldsymbol{Z}_{1}+\boldsymbol{X}_{0}-\boldsymbol{Z}_{1}\cdot\boldsymbol{X}_{0})\label{eq:ControlNot-polyn}
\end{equation}
this expression is derived from (\ref{eq:Cz-polynomial}) using the
transformation $\boldsymbol{X}_{0}=\boldsymbol{H}_{0}\cdot\boldsymbol{Z}_{0}\cdot\boldsymbol{H}_{0}$
, with $\boldsymbol{H}_{0}=\mathbf{I}\otimes\boldsymbol{H}$ and where
$\boldsymbol{H}$ is the Hadamard gate defined in (\ref{eq:Z-X-H}).

In the preceding section it has been emphasized that one can consider
different basis to define the system and the respective seed operator.
Some important properties about projection operators (not necessarily
commuting) have to outlined: 
\begin{lyxlist}{00.00.0000}
\item [{(i)}] The Kronecker product of two projection operators is also
a projection operator. 
\item [{(ii)}] If projection operators are rank-$1$ (a single eigenvalue
is $1$ all the others are $0$) then their Kronecker product is also
a rank-$1$ projection operator. 
\end{lyxlist}
So for example one can use the computational basis ($\boldsymbol{Z}$
eigenbasis) for one qubit and the $\boldsymbol{X}$ eigenbasis for
the other qubit. This is used for expressing the operator form of
the controlled-not $\boldsymbol{C}$ gate as function of the projection
operators. Taking the analogy with the $\boldsymbol{C}_{Z}$ gate
one can define the projection operator associated to the $\boldsymbol{C}$
gate as: 
\begin{equation}
\boldsymbol{\Pi}_{C}=\boldsymbol{\Pi}\otimes\boldsymbol{\Pi}_{X}=\left(\begin{array}{cc}
0 & 0\\
0 & 1
\end{array}\right)\otimes\frac{1}{2}\left(\begin{array}{cc}
1 & -1\\
-1 & 1
\end{array}\right)=\frac{1}{2}\left(\begin{array}{cccc}
0 & 0 & 0 & 0\\
0 & 0 & 0 & 0\\
0 & 0 & 1 & -1\\
0 & 0 & -1 & 1
\end{array}\right)
\end{equation}
where the projection operator on the $\boldsymbol{X}$ eigenbasis
is easily derived using $\boldsymbol{\Pi}_{X}=\boldsymbol{H}\cdot\boldsymbol{\Pi}\cdot\boldsymbol{H}$.
Then one obtains straightforwardly:

\begin{equation}
\boldsymbol{C}=\mathbf{I}-2\boldsymbol{\Pi}_{C}=(-1)^{\boldsymbol{\Pi}\otimes\boldsymbol{\Pi}_{X}}=\left(\begin{array}{cccc}
1 & 0 & 0 & 0\\
0 & 1 & 0 & 0\\
0 & 0 & 0 & 1\\
0 & 0 & 1 & 0
\end{array}\right)\label{eq:CintrilNot-projector}
\end{equation}

It is also simple to derive expression (\ref{eq:ControlNot-polyn})
from (\ref{eq:CintrilNot-projector}).

The same method can be applied to build the doubly-controlled not-gate
(Toffoli gate), named here $\boldsymbol{TO}$. One starts again with
the analogy with conjunction and notices that the gate uses 3 qubits
in a $8$-dimensional space, the compound state in the computational
basis is $|xyz>$. In the same way as before one can define here a
doubly-controlled-Z gate $\boldsymbol{C}_{\boldsymbol{C}_{Z}}$ :
\begin{equation}
\boldsymbol{C}_{\boldsymbol{C}_{Z}}=\mathbf{I}-2(\boldsymbol{\Pi}\otimes\boldsymbol{\Pi}\otimes\boldsymbol{\Pi})=(-1)^{\boldsymbol{\Pi}\otimes\boldsymbol{\Pi}\otimes\boldsymbol{\Pi}}=diag_{z}\left(1,1,1,1,1,1,1,-1\right)\:,\:\boldsymbol{C}_{\boldsymbol{C}_{Z}}|xyz>=(-1)^{xyz}|xyz>\label{eq:CCz-projector}
\end{equation}

Then the $\boldsymbol{TO}$ gate can be found by the same method as
for the $\boldsymbol{C}$ gate. The polynomial expression is easily
calculated giving:

\begin{eqnarray}
\boldsymbol{TO} & = & \boldsymbol{H}_{0}\cdot\boldsymbol{C}_{\boldsymbol{C}_{Z}}\cdot\boldsymbol{H}_{0}=\mathbf{I}-2(\boldsymbol{\Pi}\otimes\boldsymbol{\Pi}\otimes\boldsymbol{\Pi}_{X})=(-1)^{\boldsymbol{\Pi}\otimes\boldsymbol{\Pi}\otimes\boldsymbol{\Pi}_{X}}\label{eq:TO-projector}\\
 & = & \frac{1}{2}.\left(\mathbf{I}+\boldsymbol{Z}_{2}+\boldsymbol{C}-\boldsymbol{Z}_{2}\cdot\boldsymbol{C}\right)\label{eq:TO-polynomial-C}
\end{eqnarray}

The $\boldsymbol{C}$ gate can be expanded in equation (\ref{eq:TO-polynomial-C})
using equation (\ref{eq:ControlNot-polyn}) giving the alternative
expression as function of single qubit gates: 
\begin{equation}
\boldsymbol{TO}=\frac{1}{4}\left(3\mathbf{I}+\boldsymbol{Z}_{2}+\boldsymbol{Z}_{1}+\boldsymbol{X}_{0}-\boldsymbol{Z}_{2}\cdot\boldsymbol{Z}_{1}-\boldsymbol{Z}_{2}\cdot\boldsymbol{X}_{0}-\boldsymbol{Z}_{1}\cdot\boldsymbol{X}_{0}+\boldsymbol{Z}_{2}\cdot\boldsymbol{Z}_{1}\cdot\boldsymbol{X}_{0}\right)
\end{equation}

In quantum circuits it is practically difficult to realize the sum
of operators and one prefers, if it is possible, to use a product
form representing the same operator. Using the self-inverse symmetry
of the above operators it is a standard procedure to make this transformation
\cite{key-8} using (\ref{eq:General-Householder}), a simple example
is given by the transform (\ref{eq:Householder-Z}). The $\boldsymbol{C}_{Z}$
gate polynomial expression in equation (\ref{eq:Cz-polynomial}) can
be transformed using (\ref{eq:General-Householder}), (\ref{eq:Cz-polynomial})
and (\ref{eq:Cz-projector}) in the following way:

\begin{eqnarray}
\boldsymbol{C}_{Z} & = & (-1)^{\boldsymbol{\Pi}_{\boldsymbol{C}_{Z}}}=e^{i\pi\boldsymbol{\Pi}_{\boldsymbol{C}_{Z}}}=e^{i\frac{\pi}{2}}e^{-i\frac{\pi}{2}\left[\frac{1}{2}(\mathbf{I}+\boldsymbol{Z}_{1}+\boldsymbol{Z}_{0}-\boldsymbol{Z}_{1}\cdot\boldsymbol{Z}_{0})\right]}\nonumber \\
 & = & e^{+i\frac{\pi}{4}}\,\left(e^{-i\frac{\pi}{4}\boldsymbol{Z}_{1}}\right)\cdot\left(e^{-i\frac{\pi}{4}\boldsymbol{Z}_{0}}\right)\cdot\left(e^{+i\frac{\pi}{4}\boldsymbol{Z}_{1}\cdot\boldsymbol{Z}_{0}}\right)\label{eq:Cz-factoriz}
\end{eqnarray}

The factorization of the exponential operators is allowed because
all the argument operators in the exponential commute, the order of
the multiplication can thus be interchanged.

The same method can be used to obtain an expression for the controlled-not
$\boldsymbol{C}$ gate, one just replaces $\boldsymbol{Z}_{0}$ by
$\boldsymbol{X}_{0}$.

In the same way this leads to a new factorized expression of the Toffoli
gate\textit{ }$\boldsymbol{TO}$: 
\begin{eqnarray}
\boldsymbol{TO} & = & e^{+i\frac{\pi}{8}}\cdot\left(e^{-i\frac{\pi}{8}\boldsymbol{Z}_{2}}\right)\cdot\left(e^{-i\frac{\pi}{8}\boldsymbol{Z}_{1}}\right)\cdot\left(e^{-i\frac{\pi}{8}\boldsymbol{X}_{0}}\right)\cdot\nonumber \\
 &  & \cdot\left(e^{+i\frac{\pi}{8}\boldsymbol{Z}_{2}\cdot\boldsymbol{X}_{0}}\right)\cdot\left(e^{+i\frac{\pi}{8}\boldsymbol{Z}_{2}\cdot\boldsymbol{Z}_{1}}\right)\cdot\left(e^{+i\frac{\pi}{8}\boldsymbol{Z}_{1}\cdot\boldsymbol{X}_{0}}\right)\cdot\left(e^{-i\frac{\pi}{8}\boldsymbol{Z}_{2}\cdot\boldsymbol{Z}_{1}\cdot\boldsymbol{X}_{0}}\right)\label{eq:Toffoli-factioriz}
\end{eqnarray}
this formulation shows also that it is easy to scale up the gates
for example with a Toffoli-$4$ gate using three control bits on a
$4$ qubit state $|xyzw>$.

\subsection{Correspondence with recent T-gate based methods}

There has been much interest recently for developing general methods
for synthesizing quantum gates based on polynomial methods \cite{key-844,key-85,key-86}.
The decomposition of arbitrary gates into Clifford and $\boldsymbol{T}$-set
gates is an important problem. It is often desirable to find decompositions
that are optimal with respect to a given cost function. The exact
cost function used is application dependent; some possibilities are:
the total number of gates; the total number of $\boldsymbol{T}$ gates;
the circuit depth and/or the number of ancillas used.

The single-qubit non-Clifford gate $\boldsymbol{T}$ and Clifford
gate $\boldsymbol{S}$ are derived from the $\boldsymbol{Z}$ gate
and are expressed in the computational basis in their matrix form:

\begin{equation}
\boldsymbol{T}=\boldsymbol{Z}^{\frac{1}{4}}=\left(\begin{array}{cc}
+1 & 0\\
0 & e^{+i\frac{\pi}{4}}
\end{array}\right)\qquad,\qquad\boldsymbol{S}=\boldsymbol{Z}^{\frac{1}{2}}=\left(\begin{array}{cc}
+1 & 0\\
0 & e^{+i\frac{\pi}{2}}
\end{array}\right)
\end{equation}
as stated above these operators are diagonal in the computational
basis.

Considering the preceding discussion the seed operator $\boldsymbol{\Lambda}$
for a $\boldsymbol{T}$-set is the $\boldsymbol{T}$-gate itself with
eigenvalues $\left\{ +1,\omega\right\} $, naming $\omega=e^{+i\frac{\pi}{4}}$.

The action of the $\boldsymbol{T}$-gate on a qubit in the computational
basis is: $\boldsymbol{T}\,|x>=\omega^{x}\,|x>$. One can also define
the conjugate transpose gate $\boldsymbol{T}^{\dagger}|x>=(\omega^{\dagger})^{x}\,|x>=e^{-i\frac{\pi}{4}x}|x>=\omega^{-x}\,|x>$.

Using the two following arithmetic expressions of exclusive disjunction,
($XOR$, $\oplus$), for $2$ and $3$ Boolean arguments \cite{key-2}:

\begin{equation}
x\oplus y\oplus z=x+y+z-2xy-2xz-2yz+4xyz\quad,\qquad x\oplus y=x+y-2xy
\end{equation}
where the second member is an inclusion-exclusion-like form. Combining
the expressions gives:

\begin{equation}
4xyz=x+y+z-x\oplus y-x\oplus z-y\oplus z+x\oplus y\oplus z\label{eq:xyz-xor}
\end{equation}
this last expression gives a method for building more complex gates
using only $\boldsymbol{T}$ and $\boldsymbol{T}^{\dagger}$ gates
as shown by Peter Selinger in \cite{key-85}.

Starting again with the double controlled-Z gate $\boldsymbol{C}_{\boldsymbol{C}_{Z}}$
one uses the fact, see (\ref{eq:Householder-Z}), that $\boldsymbol{T}^{4}=\boldsymbol{Z}$
and thus $\boldsymbol{T}=\boldsymbol{Z}^{\frac{1}{4}}=e^{i\frac{\pi}{8}}e^{-i\frac{\pi}{8}\boldsymbol{Z}}$.
One defines the $3$-qubit operators: $\boldsymbol{T}_{0}=\mathbf{I}\otimes\mathbf{I}\otimes\boldsymbol{T}$
, $\boldsymbol{T}_{1}=\mathbf{I}\otimes\boldsymbol{T}\otimes\mathbf{I}$
and $\boldsymbol{T}_{2}=\boldsymbol{T}\otimes\mathbf{I}\otimes\mathbf{I}$.
Using (\ref{eq:xyz-xor}) and the Reed-Muller methods discussed in
section \ref{subsec:Fourier-transform-of} one can express the $\boldsymbol{C}_{\boldsymbol{C}_{Z}}$
operator using the action on the $3$-qubit state $|xyz>$ as defined
in (\ref{eq:CCz-projector}) and the eigenvalue relation $(-1)^{x}=\omega^{4x}$:

\begin{equation}
\boldsymbol{C}_{\boldsymbol{C}_{Z}}=\boldsymbol{T}_{0}\cdot\boldsymbol{T}_{1}\cdot\boldsymbol{T}_{2}\cdot(\boldsymbol{T}_{x\oplus y}^{[3]})^{\dagger}\cdot(\boldsymbol{T}_{x\oplus z}^{[3]})^{\dagger}\cdot(\boldsymbol{T}_{y\oplus z}^{[3]})^{\dagger}\cdot(\boldsymbol{T}_{x\oplus y\oplus z}^{[3]})\label{eq:CCz-Selinger}
\end{equation}
the operators corresponding to exclusive disjunction $\oplus$ in
(\ref{eq:CCz-Selinger}) can be easily obtained using the Eigenlogic
interpretation: they are diagonal operators where the diagonal elements
are the truth values of the corresponding logical connective using
the $\boldsymbol{T}$ operator alphabet $\left\{ +1,\omega\right\} $.
For example explicitly:

\begin{equation}
\boldsymbol{T}_{x\oplus y}^{[3]}=\boldsymbol{T}_{x\oplus y}^{[2]}\otimes\mathbf{I}=diag_{z}(1,\omega,\omega,1)\otimes\mathbf{I}=diag_{z}(1,1,\omega,\omega,\omega,\omega,1,1)
\end{equation}

It can be shown, again because $\boldsymbol{T}=\boldsymbol{Z}^{\frac{1}{4}}=e^{+i\frac{\pi}{8}}e^{-i\frac{\pi}{8}\boldsymbol{Z}}$,
that the Toffoli gate $\boldsymbol{TO}=\boldsymbol{H}_{0}\cdot\boldsymbol{C}_{\boldsymbol{C}_{Z}}\cdot\boldsymbol{H}_{0}$
obtained using (\ref{eq:CCz-Selinger}) is equivalent to expression
(\ref{eq:Toffoli-factioriz}). The operator given in (\ref{eq:CCz-Selinger})
can be explicitly designed using the methods described in \cite{key-85}.

An alternative polynomial expression can be found directly by the
interpolation method. The idea is that because $\boldsymbol{T}$ and
$\boldsymbol{Z}$ commute and are not degenerate they have the same
eigenvectors and thus one can use the same projection operators which
are $\boldsymbol{\Pi}$ and its complement $\mathbf{I}-\boldsymbol{\Pi}$
(see equation (\ref{eq:Project-Z})). The expression of the double
controlled-Z gate $\boldsymbol{C}_{\boldsymbol{C}_{Z}}$ as a function
of $\boldsymbol{\Pi}$ was already calculated in (\ref{eq:CCz-projector}),
now one just has to express the projection operator $\boldsymbol{\Pi}$
as function of the operator $\boldsymbol{T}$ using (\ref{eq:Lagr_projector}),
this gives:

\begin{equation}
\boldsymbol{\Pi}_{\omega}(\boldsymbol{T})=\boldsymbol{\Pi}=(\omega-1)^{-1}(\boldsymbol{T}-\mathbf{I})\qquad,\qquad\boldsymbol{\Pi}_{+1}(\boldsymbol{T})=\mathbf{I}-\boldsymbol{\Pi}=-(\omega-1)^{-1}(\boldsymbol{T}-\omega\mathbf{I})
\end{equation}
so using directly (\ref{eq:CCz-projector}) one has:

\begin{equation}
\boldsymbol{C}_{\boldsymbol{C}_{Z}}=\mathbf{I}-2(\omega-1)^{-3}\left[(\boldsymbol{T}-\mathbf{I})\otimes(\boldsymbol{T}-\mathbf{I})\otimes(\boldsymbol{T}-\mathbf{I})\right]
\end{equation}
which can also be expressed as a function of $\boldsymbol{T}_{0}$
, $\boldsymbol{T}_{1}$ and $\boldsymbol{T}_{2}$:

\begin{equation}
\boldsymbol{C}_{\boldsymbol{C}_{Z}}=\mathbf{I}+\frac{2}{(\omega-1)^{3}}\left(\mathbf{I}-\boldsymbol{T}_{2}-\boldsymbol{T}_{1}-\boldsymbol{T}_{0}+\boldsymbol{T}_{2}\cdot\boldsymbol{T}_{1}+\boldsymbol{T}_{2}\cdot\boldsymbol{T}_{0}+\boldsymbol{T}_{1}\cdot\boldsymbol{T}_{0}-\boldsymbol{T}_{2}\cdot\boldsymbol{T}_{1}\cdot\boldsymbol{T}_{0}\right)\label{eq:Ccz function T}
\end{equation}
again using the transformation method it is easy to show that (\ref{eq:Ccz function T})
is equivalent to (\ref{eq:CCz-Selinger}). The Toffoli gate $\boldsymbol{TO}$
is also straightforwardly derived by replacing $\boldsymbol{T}_{0}$
by $(\boldsymbol{H}_{0}\cdot\boldsymbol{T}_{0}\cdot\boldsymbol{H}_{0})$
in (\ref{eq:Ccz function T}) leading again to (\ref{eq:Toffoli-factioriz}).

The same method could be employed using $\boldsymbol{S}$ gates, for
example simply by replacing the $\boldsymbol{T}$ operators by the
respective $\boldsymbol{S}$ ones and $\omega=e^{+i\frac{\pi}{4}}$
by $\omega_{S}=e^{+i\frac{\pi}{2}}$ in (\ref{eq:Ccz function T}).
Also replacing $\omega$ by $-1$ and $\boldsymbol{T}$ by $\boldsymbol{Z}$
in (\ref{eq:Ccz function T}) leads to an Eigenlogic operator expression
for the three-input conjunction $\boldsymbol{C}_{\boldsymbol{C}_{Z}}=\boldsymbol{G}_{x\wedge y\wedge z}^{[3]}$
as defined in (\ref{eq:3-G-AND}).

\section{Interpolation synthesis of multivalued quantum gates}

Multi-valued logic requires a different algebraic structure than ordinary
binary-valued one. Many properties of binary logic do not support
set of values that do not have cardinality $2^{n}$. Multi-valued
logic is often used for the development of logical systems that are
more expressive than Boolean systems for reasoning \cite{key-86}.
Particularly three and four valued systems, have been of interest
with applications in digital circuits and computer science. Quantum
physics and modern theory of many-valued logic were born nearly simultaneously
\cite{key-4} in the second and third decade of the twentieth century.
The recently observed revival of interest in applying many-valued
logic to the description of quantum phenomena is closely connected
with fuzzy logic {[}3{]}. Multivalued logic is of interest to engineers
involved in various aspects of information technology. It has a long
history of use in CAD with HDL (Hardware Description Languages) for
simulation of digital circuits and their synthesis, vaqrious standards
have been established. The total number of logical connectives for
for an $m$-valued $n$-arity system is the combinatorial number $m^{m^{n}}$
\cite{key-4}. In particular for an arity-$1$ system with $3$ values
the number of connectives will be $3^{3^{1}}=27$ and for an arity-$2$
system the number of connectives will be $3^{3^{2}}=19683$. So it
is clear that by increasing the values from $2$ to $m$ the possibilities
of new connectives becomes intractable for a complete description
of a logical system, but some special connectives play important roles
and will be illustrated hereafter.

\subsection{Multivalued operators for quantum Fourier transform}

Following David Mermin in \cite{key-8} one can introduce the following
unitary operator $\boldsymbol{\Im}$ in a $2^{n}$dimension Hilbert
space. Its action on the computational basis for a $n$-qubit state
$|\boldsymbol{p}^{[n]}>$ is:

\begin{equation}
\boldsymbol{\Im}|\boldsymbol{p}^{[n]}>=e^{-\tfrac{2i\pi}{2^{n}}d_{p}}|\boldsymbol{p}^{[n]}>\qquad,\quad\left(\boldsymbol{\Im}\right)^{2^{n}}=\mathbf{I}_{2^{n}}\label{eq:F-operator-FT}
\end{equation}
where $|\boldsymbol{p}^{[n]}>=|p_{n-1},...,p_{0}>$ is the compound
quantum state and $0\leq d_{p}\leq2^{n}-1$ is the (decimal) number
corresponding to the register of the state with binary digits (bits)
$p_{i}\in\{0,1\}$. The cyclic character of this operator is also
showed in (\ref{eq:F-operator-FT}).

The unitary operator $\boldsymbol{U}_{FT}$ corresponding to the quantum
Fourier transform operation is then defined by:

\begin{equation}
\boldsymbol{U}_{FT}|\boldsymbol{q}^{[n]}>=\frac{1}{\sqrt{2^{n}}}\sum_{d_{p}=0}^{2^{n}-1}e^{-\tfrac{2i\pi}{2^{n}}d_{q}d_{p}}|\boldsymbol{p}^{[n]}>=\left(\boldsymbol{\Im}\right)^{d_{q}}\cdot\boldsymbol{H}^{\otimes n}|\boldsymbol{0}^{[n]}>\label{eq:U-QFT}
\end{equation}
where $|\boldsymbol{0}^{[n]}>$ is the vector with all qubit digits
at zero, $\boldsymbol{H}^{\otimes n}$ is the operator obtained by
the Kronecker product of the $n$ Hadamard operators (\ref{eq:Z-X-H}).

The $\boldsymbol{\Im}$ operator has a non-degenerate finite spectrum
of eigenvalues, the roots of unity, and can thus be considered as
a multivalued seed operator. An example can be illustrated where the
seed projector can be directly written as function of the interpolation
projection operators for the qubits:

\begin{equation}
\boldsymbol{\Pi}_{k}=\boldsymbol{\mathrm{I}}_{2}^{\otimes(2^{n}-1-k)}\otimes\boldsymbol{\Pi}\otimes\boldsymbol{\mathrm{I}}_{2}^{\otimes k}\quad,\qquad k\in\{0,2^{n}-1\}
\end{equation}
leading for a $4$-qubit system to the operator:

\begin{equation}
\boldsymbol{\Im}_{n=4}=e^{-\tfrac{2i\pi}{16}(8\boldsymbol{\Pi}_{3}+4\boldsymbol{\Pi}_{2}+2\boldsymbol{\Pi}_{1}+\boldsymbol{\Pi}_{0})}
\end{equation}

Here one can make a parallel with angular momentum observables which
are generators for rotations. The eigenvalues $\hbar m$ of the $z$
component angular momentum $\boldsymbol{J}_{z}$ obey the following
rules $-j\leq m\leq+j$ and the difference between successive values
is $\varDelta m=1$. $j\geq0$ is integer or half-integer and $\hbar^{2}j(j+1)$
is the eigenvalue of the associated observable $\boldsymbol{J}^{2}=\boldsymbol{J}_{x}^{2}+\boldsymbol{J}_{y}^{2}+\boldsymbol{J}_{z}^{2}$
. For cardinality $2^{n}$ one has $j^{[n]}=\tfrac{2^{n}-1}{2}$ which
is half-integer, the considered numerical system becomes:

\begin{eqnarray}
m^{[n]} & \in & \frac{1}{2}\left\{ -2^{n}+1,...,-1,+1,...,+2^{n}-1\right\} \nonumber \\
m^{[n]}+j^{[n]}=d^{[n]} & \in & \left\{ 0,1,2...,+2^{n}-1\right\} \label{eq:AM-eigenvalu-number}
\end{eqnarray}
this shows that by shifting the spectrum one can express $\boldsymbol{\Im}$
as a function of $\boldsymbol{J}_{z}$ leading to the operator expression:

\begin{equation}
\boldsymbol{\Im}=-e^{+\tfrac{i\pi}{2^{n}}}e^{-\tfrac{2i\pi}{2^{n}}\tfrac{\boldsymbol{J}_{z}}{\hbar}}
\end{equation}

The operator $\boldsymbol{\Im}$ which is a function of the physical
observable angular momentum can be thus associated with physical systems.
The operator $\boldsymbol{U}_{FT}$ according to (\ref{eq:U-QFT})
can be expressed as a function of the product of two angular momentum
operators $\boldsymbol{J}_{z}$ which could represent magnetic spin-spin
interaction Hamiltonians.

\subsection{$\left\{ +1,0,-1\right\} $ OAM system, for ternary Min and Max logical
gates}

The logical system $\{+1,0,-1\}$ has several benefits because it
approaches the two logical systems most commonly used in binary logic
$\left\{ +1,-1\right\} $ and $\left\{ 0,1\right\} $, which are special
cases of the considered ternary logic. Moreover, its values are centered
in zero, thus assuring a simplification of the results and interesting
properties of symmetry.

Orbital angular momentum (OAM) is characterized by two quantum numbers:
$\ell$ the orbital number and $m$ the magnetic number. The rules
are: $\ell\geq0$ is an integer and $-\ell\leq m\leq\ell$. The matrix
form of the $z$-component orbital angular momentum observable for
$\ell=1$ is:

\begin{equation}
\boldsymbol{L}_{z}=\hbar\boldsymbol{\varLambda}=\hbar\left(\begin{array}{ccc}
+1 & 0 & 0\\
0 & 0 & 0\\
0 & 0 & -1
\end{array}\right)\label{eq:Lz}
\end{equation}
the three eigenvalues $\{+1,0,-1\}$ are here considered the logical
truth values.

One can now express the ternary logical observables as developments
over the rank-$1$ projection operators spanning the vector space:
$\boldsymbol{\Pi}_{+1}$, $\boldsymbol{\Pi}_{0}$ and $\boldsymbol{\Pi}_{-1}$.
These operators are explicitly calculated using equation (\ref{eq:Lagr_projector})
and correspond to the density matrices of the three eigenstates $|+1>$
, $|0>$ and $|-1>$ of $\boldsymbol{L}_{z}$, defining a qutrit.
The projection operators function of the seed operator $\boldsymbol{\varLambda}$
are given by:

\begin{equation}
\boldsymbol{\Pi}_{+1}=\frac{1}{2}\boldsymbol{\varLambda}\left(\boldsymbol{\varLambda}+\boldsymbol{I}\right)\quad,\qquad\boldsymbol{\Pi}_{0}=\boldsymbol{I}-\boldsymbol{\varLambda}^{2}\quad,\qquad\boldsymbol{\Pi}_{-1}=\frac{1}{2}\boldsymbol{\varLambda}\left(\boldsymbol{\varLambda}-\boldsymbol{I}\right)\label{eq:Project-3v}
\end{equation}

All arity-one logical operators $\boldsymbol{F}(\boldsymbol{\varLambda})$
can then be derived using the development (\ref{eq:1-arity-Logc-funct-dev}).

When considering an arity-$2$ , $3$-valued system, the operators
are represented by $9\times9$ matrices. The dictators, $\boldsymbol{U}$
and $\boldsymbol{V}$, are then:

\begin{equation}
\boldsymbol{U}=\boldsymbol{\varLambda}\otimes\boldsymbol{I}\qquad,\qquad\boldsymbol{V}=\boldsymbol{I}\otimes\boldsymbol{\varLambda}\qquad,\qquad\boldsymbol{U}\cdot\boldsymbol{V}=\boldsymbol{\varLambda}\otimes\boldsymbol{\varLambda}\label{eq:Dictat-3v 2a}
\end{equation}

In ternary logic, popular connectives are $\mathrm{Min}$ and $\mathrm{Max}$,
defined by their truth-value maps in Table \ref{tab:Min Max}.

Using (\ref{eq:Project-3v}) and (\ref{eq:Dictat-3v 2a}) and logical
reduction rules (due to the completeness of the projection operators)
one obtains the following observables: 
\begin{equation}
\left\{ \begin{array}{l}
{\displaystyle \boldsymbol{\mathrm{Min}}(\boldsymbol{U},\boldsymbol{V})\,=\,\frac{1}{2}\,\big(\boldsymbol{U}+\boldsymbol{V}+\boldsymbol{U}^{2}+\boldsymbol{V}^{2}-\boldsymbol{U}\cdot\boldsymbol{V}-\boldsymbol{U}^{2}\cdot\boldsymbol{V}^{2}\big)=diag(+1,+1+1,+1,0,0,+1,0,-1)}\\
{\displaystyle \vspace{-.2cm}~}\\
{\displaystyle \boldsymbol{\mathrm{Max}}(\boldsymbol{U},\boldsymbol{V})\,=\,\frac{1}{2}\,\big(\boldsymbol{U}+\boldsymbol{V}-\boldsymbol{U}^{2}-\boldsymbol{V}^{2}+\boldsymbol{U}\cdot\boldsymbol{V}+\boldsymbol{U}^{2}\cdot\boldsymbol{V}^{2}\big)}=diag(+1,0,-1,0,0,-1,-1,-1,-1)
\end{array}\right.\label{minmax}
\end{equation}

\begin{table}[h]
\centering{}%
\begin{tabular}{|>{\centering}p{2.4cm}||>{\centering}p{0.7cm}||>{\centering}p{0.7cm}||>{\centering}p{0.7cm}|}
\hline 
Min : $U\;\setminus\;V$  & $+1$  & $0$  & $-1$\tabularnewline
\hline 
\hline 
$+1$  & $+1$  & $+1$  & $+1$\tabularnewline
\hline 
\hline 
$0$  & $+1$  & $0$  & $0$\tabularnewline
\hline 
\hline 
$-1$  & $+1$  & $0$  & $-1$\tabularnewline
\hline 
\end{tabular}\quad{}%
\begin{tabular}{|>{\centering}p{2.4cm}||>{\centering}p{0.7cm}||>{\centering}p{0.7cm}||>{\centering}p{0.7cm}|}
\hline 
Max : $U\;\setminus\;V$  & $+1$  & $0$  & $-1$\tabularnewline
\hline 
\hline 
$+1$  & $+1$  & $0$  & $-1$\tabularnewline
\hline 
\hline 
$0$  & $0$  & $0$  & $-1$\tabularnewline
\hline 
\hline 
$-1$  & $-1$  & $-$1  & $-1$\tabularnewline
\hline 
\end{tabular}\caption{\label{tab:Min Max}The Min and Max tables for a three-valued two-argument
logic}
\end{table}

\subsection{Operators for a qutrit balanced half-adder $\left\{ -1,0,+1\right\} $}

The adder is one of the fundamental elements in digital electronics.
One can, by the means of multivalued logic, improve the performances
of this circuit by removing the delays caused by the propagation of
the carry bit.

Considering a balanced half-adder, the structure is simplified because
the ternary logic values match the notation for balanced ternary digits.
The truth table for a balanced ternary half adder is given in Table
\ref{tab:Half adder} \cite{key-9}, where $\mathrm{S_{i}}$ represents
the sum and $\mathrm{C_{i+1}}$ the carry-out logical connectives.
The inputs for the half-adder are the addend value $\mathrm{A}$ and
the carry-in value $\mathrm{C_{i}}$.

\begin{table}[h]
\centering{}%
\begin{tabular}{|>{\centering}p{2.4cm}||>{\centering}p{0.7cm}||>{\centering}p{0.7cm}||>{\centering}p{0.7cm}|}
\hline 
$\mathrm{S_{i}:}$ $\mathrm{A\:}\setminus\:\mathrm{C_{i}}$  & $-1$  & $0$  & $+1$\tabularnewline
\hline 
\hline 
$-1$  & $+1$  & $-1$  & $+0$\tabularnewline
\hline 
\hline 
$0$  & $-1$  & $0$  & $+1$\tabularnewline
\hline 
\hline 
$\text{+}1$  & $0$  & $+1$  & $-1$\tabularnewline
\hline 
\end{tabular}\quad{}%
\begin{tabular}{|>{\centering}p{2.4cm}||>{\centering}p{0.7cm}||>{\centering}p{0.7cm}||>{\centering}p{0.7cm}|}
\hline 
$\mathrm{C_{i+1}}:$ $\mathrm{A\:}\setminus\:\mathrm{C_{i}}$  & $-1$  & $0$  & $+1$\tabularnewline
\hline 
\hline 
$-1$  & $-1$  & $0$  & $0$\tabularnewline
\hline 
\hline 
$0$  & $0$  & $0$  & $0$\tabularnewline
\hline 
\hline 
$+1$  & $0$  & $0$  & $+1$\tabularnewline
\hline 
\end{tabular}\caption{\label{tab:Half adder}The Half-Adder tables for a balanced three-valued
two-argument logic}
\end{table}

The seed operator is reversed from the one given above: $\boldsymbol{\varLambda}_{ha}=diag(-1,0,+1)$.
The inputs are represented by the addend operator $\boldsymbol{A}=\boldsymbol{\varLambda}_{ha}\otimes\boldsymbol{I}$
and the carry-in operator $\boldsymbol{C}_{i}=\boldsymbol{I}\otimes\boldsymbol{\varLambda}_{ha}$.
The output operators are the sum $\boldsymbol{S}_{i}$ and the carry-out
$\boldsymbol{C}_{i+1}$ operators, their spectrum correspond to the
respective truth values and is explicitly given in (\ref{Half-Adder}).
The corresponding logical observables are derived directly using the
truth tables and the respective projection operators (same method
as for the OAM case using (\ref{eq:Project-3v})):

\begin{equation}
\left\{ \begin{array}{l}
{\displaystyle \boldsymbol{S}_{i}(\boldsymbol{A},\boldsymbol{C}_{i})\,=\,\boldsymbol{A}+\boldsymbol{C}_{i}-\frac{3}{2}\,\boldsymbol{A}^{2}\cdot\boldsymbol{C}_{i}-\frac{3}{2}\,\boldsymbol{A}\cdot\boldsymbol{C}_{i}^{2}=diag(+1,-1,0,-1,0,+1,0,+1,-1)}\\
{\displaystyle \vspace{-.2cm}~}\\
{\displaystyle \boldsymbol{C}_{i+1}(\boldsymbol{A},\boldsymbol{C}_{i})\,=\,\frac{1}{2}\,\big(\boldsymbol{A}\cdot\boldsymbol{C}_{i}^{2}+\boldsymbol{A}^{2}\cdot\boldsymbol{C}_{i}\big)}=diag(-1,0,0,0,0,0,0,0,+1)
\end{array}\right.\label{Half-Adder}
\end{equation}

The comparison of the truth tables using ternary logic with the binary
ones shows that the sum function $\mathrm{S_{i}}$ corresponds here
to the modulo-3 sum function, and the carry function $\mathrm{C_{i+1}}$
is the consensus function. Therefore, the implementation of a balanced
ternary half-adder is natural. The half-adder will either increment
or decrement, depending on whether $\mathrm{C_{0}}$ is $+1$ or $-1$
, while the binary equivalent can only increment.

\section{Conclusion}

A general method has been presented for the design of logical quantum
gates. It uses matrix interpolation inspired from classical multivariate
methods. The interesting property is that a unique seed operator generates
the entire logical family of operators for a given $m$-valued $n$-arity
system. When considering the binary alphabet $\{+1,-1\}$ the gates
are the quantum equivalent of the Fourier transform of Boolean functions.
The method proposes a new expression of the Toffoli gate which can
be understood as a 3-argument conjunction in the Eigenlogic interpretation.
For multivalued logic quantum gates can be derived for different alphabets.
The correspondence with quantum angular momentum leads to physical
realizable gates. Applications have been presented for quantum Fourier
transform gates, Min-Max and half-adder logical operations.

This opens a new perspective for quantum computation because several
of the Eigenlogic operators \cite{key-2,key-3} turn out to be well-known
quantum gates. This shows an operational correspondence between quantum
control logic (Deutsch's paradigm \cite{key-1}) and ordinary propositional
logic. In Eigenlogic measurements on logical operators give the truth
values of the corresponding logical connective. How could these measurement
be exploited in a quantum circuit? Quantum tomography inspired techniques
could be used.

At first sight the methods discussed here could be viewed as ``classical''
because of the identification of Eigenlogic with propositional logic.
But when considering quantum states,\textit{ }that are not eigenvectors,
the measurement outcomes are governed by the probabilistic quantum
Born rule, and interpretable results are then the mean values, this
leads to a fuzzy logic interpretation as outlined in \cite{key-3}.

\section*{Acknowledgements }

We would like to thank Benoît Valiron from CentraleSupélec and LRI
(Laboratoire de Recherche en Informatique), Gif-sur-Yvette (FR) for
fruitful discussions during an ongoing academic project on quantum
programming and for having pointed out the work using $T$ gates.
We are also very grateful to Francesco Galofaro, from Politecnico
Milano (IT) and Free University of Bolzano (IT) for his pertinent
advices on semantics and logic.

We much appreciated the feedback from the \textit{Quantum Interaction}
community these last years, the foundational aspects related to this
work were presented at the QI-2016 conference in San Francisco and
we wish to thank particularly José Acacio da Barros of San Francisco
State University (CA, ,USA), Ehtibar Dzhafarov of Purdue University
(IN, USA), Andrei Khrennikov of Linnaeus University (SWE), Dominic
Widdows from Microsoft Bing Bellevue (WA, USA), Peter Wittek from
ICFO (Barcelona ESP) and Keith van Rijsbergen from University of Glasgow
(UK).


\begin{thebibliography}{10}
\bibitem{key-1} David Deutsch, ``Quantum theory, the Church-Turing
principle and the universal quantum computer\textquotedblright , Proc.
Royal Soc. A. 400 (1818): pp. 97\textendash 117, 1985.

\bibitem{key-2} Zeno Toffano, ``Eigenlogic in the spirit of George
Boole\textquotedblright , ArXiv:1512.06632, 2015.

\bibitem{key-3} François Dubois and Zeno Toffano, ``Eigenlogic:
a Quantum View for Multiple-Valued and Fuzzy Systems\textquotedblright ,
Quantum Interaction. QI 2016. Lecture Notes in Computer Science, vol
10106. Springer, pp. 239-251, 2017, arXiv:1607.03509 {[}quant-ph{]}.

\bibitem{key-30} Pierre Cartier, ``A mad day's work: from Grothendieck
to Connes and Kontsevich The evolution of concepts of space and symmetry\textquotedblright ,
Journal: Bull. Amer. Math. Soc. 38 (2001), 389-408.

\bibitem{key-31} John von Neumann, ``Mathematische Grundlagen der
Quantenmechanik'', in ``Grundlehren der mathematischen Wissenschaften'',
volume Bd. 38. (Springer, Berlin, 1932) 106. ``Mathematical Foundations
of Quantum Mechanics'', Investigations in Physics, vol. 2, Princeton
University Press, Princeton, 1955.

\bibitem{key-32} Garret Birkhoff, John von Neumann, ``The Logic
of Quantum Mechanics\textquotedblright . The Annals of Mathematics,
2nd Ser., 37 (4), 823-843, 1936.

\bibitem{key-33} Jeffrey Bub, ``Quantum computation from a quantum
logical perspective\textquotedblright , arXiv:quant-ph/0605243.

\bibitem{key-34} Mingsheng Ying, ``A theory of computation based
on quantum logic (I)\textquotedblright , Theoretical Computer Science,
Volume 344, Issues 2\textendash 3, Pages 134-207, 2005.

\bibitem{key-4} Emil Post, ``Introduction to a General theory of
Elementary Propositions \textquotedblright , American Journal of Mathematics
43: 163\textendash 185, 1921.

\bibitem{key-6} Svetlana N. Yanushkevich, Vlad P. Shmerko, ``Introduction
to Logic Design'', CRC Press, 2008.

\bibitem{key-7} Noboru Kikuchi, ``Finite Element Methods in Mechanics'',
Cambridge University Press, 1986.

\bibitem{key-8} David Mermin, ``Quantum Computer Science. An Introduction'',
Cambridge University Press, 2007.

\bibitem{key-81} George Boole, ``The Mathematical Analysis of Logic.
Being an Essay To a Calculus of Deductive Reasoning'', reissued Ed.
Forgotten Books ISBN 978-1444006642-9, 1847.

\bibitem{key-82} Ryan O'Donnell, ``Analysis of Boolean Functions'',
Cambridge University Press, 2014.

\bibitem{key-83} Ashley Montanaro, Tobias J. Osborne, ``Quantum
boolean functions\textquotedblright , Chicago Journal of Theoretical
Computer Science, vol 2010, n\textdegree 1, 2010, arXiv:0810.2435
{[}quant-ph{]}

\bibitem{key-84} D. E. Muller, ``Application of Boolean Algebra
to Switching Circuit Design and to Error Detection\textquotedblright ,
IRE Transactions on Electronic Computers, 3:6\textendash 12, 1954.

\bibitem{key-844} Matthew Amy, Michele Mosca, ``T-count optimization
and Reed-Muller codes\textquotedblright , arXiv:1601.07363.

\bibitem{key-85} Peter Selinger, ``Quantum circuits of T-depth one'',
Phys. Rev. A 87, 042302, 2013.

\bibitem{key-86} D. Michael Miller, Mitchell Aaron Thornton, ``Multiple
Valued Logic: Concepts and Representations\textquotedblright , Morgan
\& Claypool Publishers, 2008.

\bibitem{key-9} Douglas W. Jones, ``Fast Ternary Addition'', http://www.cs.uiowa.edu/\textasciitilde{}jones/ternary/
, University of Iowa Dep. of Computer Science, 2013.
\end{thebibliography}
\end{document}